\documentclass{ws-ijbc}
\usepackage{ws-rotating}     

\newcommand{\documentdate}{29 November 2010}

\begin{document}

\thispagestyle{empty}
\includegraphics[height=3.5cm]{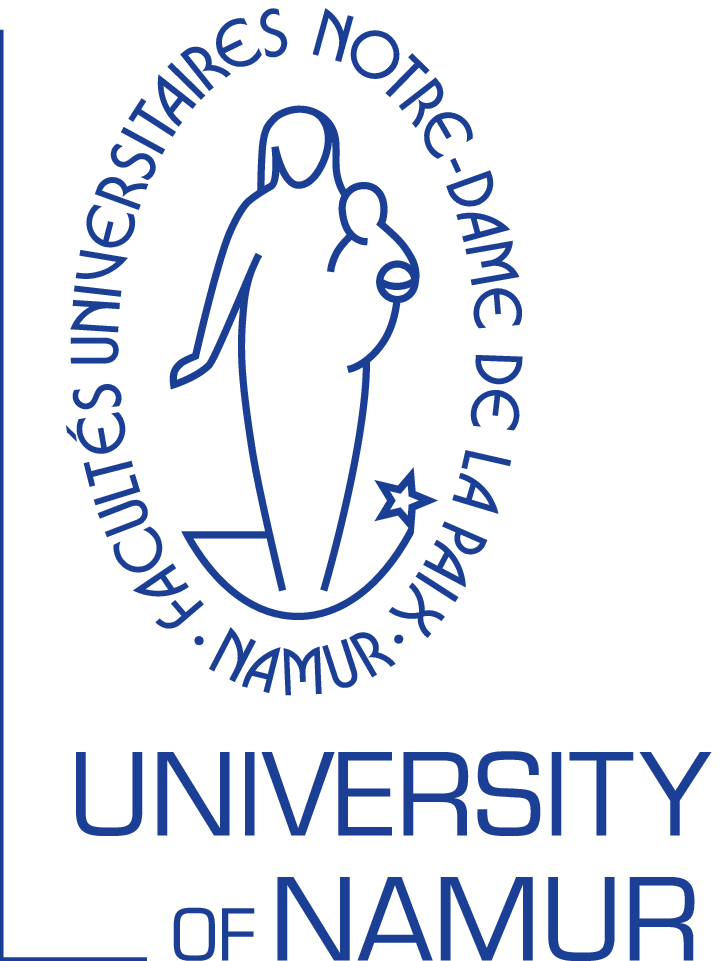}

\vspace*{2cm}
\hspace*{1.3cm}
\fbox{\rule[-3cm]{0cm}{6cm}\begin{minipage}[c]{12cm}

\begin{center}
On the use of the MEGNO indicator with the Global Symplectic
Integrator\\
\mbox{}\\
by Ch. HUBAUX, A.-S. LIBERT and T. CARLETTI\\
\mbox{}\\
Report naXys-06-2010 \hspace*{20mm} \documentdate 
\end{center}
\end{minipage}
}

\vspace{2cm}
\begin{center}
\includegraphics[height=3.5cm]{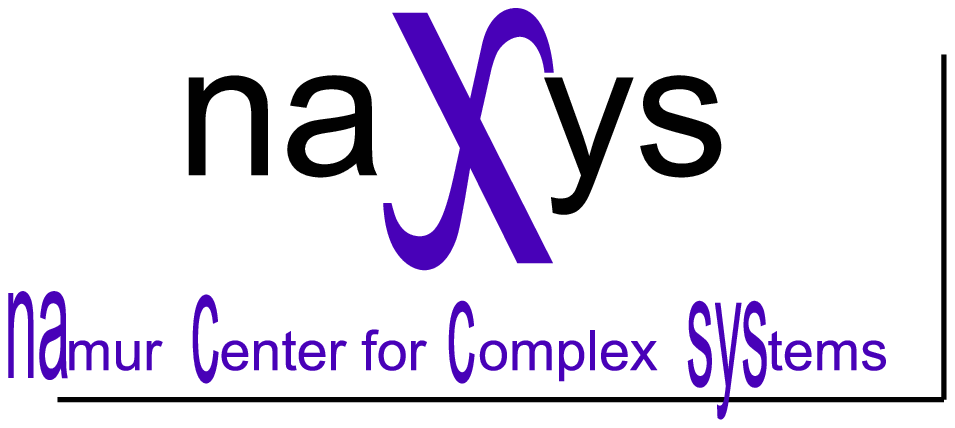}
{\Large \bf Put the naXys logo here}

\vspace{2cm}
{\Large \bf Namur Center for Complex Systems}

{\large
University of Namur\\
8, rempart de la vierge, B5000 Namur (Belgium)\\*[2ex]
{\tt http://www.naxys.be}}

\end{center}

\newpage
\setcounter{page}{1}

\catchline{}{}{}{}{} 

\markboth{Ch. HUBAUX et al.}{On the use of the MEGNO indicator with the Global Symplectic Integrator}

\title{ (This paper is for the Special Issue edited by \\ Prof. Gregoire Nicolis , Prof. Marko Robnik, Dr. Vassilis Rothos and Dr. Haris Skokos) \\ On the use of the MEGNO indicator with the Global Symplectic Integrator}

\author{Ch. HUBAUX, A.-S. LIBERT and T. CARLETTI }
\address{naXys, Namur Center for Complex Systems \\ Department of Mathematics, University of Namur \\
8 rempart de la vierge, B5000 Belgium\\
\{charles.hubaux, anne-sophie.libert, timoteo.carletti\}@fundp.ac.be}

\maketitle

\begin{history}
\received{(to be inserted by publisher)}
\end{history}

\begin{abstract}
{To distinguish between regular and chaotic orbits in Hamiltonian systems, the
  \emph{Global Symplectic Integrator} (GSI) has been
  introduced~\cite{Libert10}, based on the symplectic integration of both
  Hamiltonian equations of motion and variational equations. In the present
  contribution, we show how to compute efficiently the MEGNO indicator jointly
  with the GSI. Moreover, we discuss the choice of symplectic integrator,
  in fact we point out that a particular attention
  has to be paid to the structure of the Hamiltonian system associated to the
  variational equations. The 
  performances of our method is illustrated through the study of the Arnold
  diffusion problem.} 
\end{abstract}


\newcommand{\R}{\mathbb{R}}
\newcommand{\T}{\mathbb{T}}
\newcommand{\Z}{\mathbb{Z}}

\section{Introduction}

Several detection techniques exist in order to study the regular or chaotic
behavior of orbits of Hamiltonian systems. 
All \emph{Lyapunov}-like methods are based on the resolution of the so-called
variational equations giving the evolution of deviation vectors associated to
a given orbit.  
These are essentially the FLI \cite{Froeschle97}, MEGNO \cite{Cincotta03},
SALI \cite{Skokos01} and GALI \cite{Skokos07} methods.  
In \cite{Libert10}, we have introduced the \emph{Global Symplectic Integrator}
(for short the GSI), a method allowing to solve both
Hamiltonian equations of motion and variational equations using a totally
  symplectic integration scheme. 
Based on a comparison of the GSI
with non-symplectic integration schemes, it was 
clearly shown that the GSI was more
accurate in the detection of regular and chaotic orbits, using the SALI
  chaos indicator, and less time-consuming than non-symplectic methods. Such
    a kind of symplectic integration scheme for the variational equations has
    also been recently and independently proposed in \cite{Skokos10}. 

The purpose of this work is to show that other chaos detection techniques, for
instance the MEGNO, can be used jointly with the GSI and their efficiency
  thus improved.  
The computation of the MEGNO needs to study the time evolution of a single
deviation vector. 
This advantage, with respect to the SALI that needs the knowledge of the
  evolution of two deviation vectors, becomes critical when considering a
large amount of orbits on long time spans.  
Moreover, the use of the GSI does not rely on the use of a specific symplectic
integrator scheme. Hence, our second aim is to discuss and compare different
symplectic integrators.  
It turns out that methods adapted to the structure of the Hamiltonian equations of motion are
not necessarily suited for the associated variational equations. 

While the study of the dynamics of two well-known Hamiltonian systems,
H\'enon-Heiles and the restricted three-body problem, have been addressed in
\cite{Libert10}, we hereby propose to use the GSI to improve the detection of
slow diffusion in Hamiltonian systems, the so-called Arnold diffusion,
analyzed according to the model proposed in~\cite{Lega03}.

The organization of the paper is as follows.
In Section~\ref{sec:gsi}, we briefly {summarize} the GSI method, and we review
two classes of symplectic integrators. Section~\ref{sec:megno} is devoted to
the definition of the MEGNO {indicator} and the algorithm used to compute it. 
{An application to the Arnold diffusion problem} is then presented in
Section~\ref{sec:arndiff}, for which the the GSI method is compared to a
non-symplectic scheme. Finally, in Section~\ref{sec:concl} we sum up and
  draw our conclusions.

\section{Global Symplectic Integrator}
\label{sec:gsi}

\subsection{Method}

Let us consider an autonomous Hamiltonian system with $N$ degrees of freedom ${\mathcal H}(\bf p,\bf q)$ where $\bf p,\bf q\in \R^N$ are the momenta and variables vectors. The Hamiltonian vector field may be written as 
\begin{equation} 
 \dot {\bf x} = \mathcal J {\bf \nabla}_{\bf x} \mathcal H = {\bf W}({\bf
   x})\, ,  \label{HVectField}
\end{equation}
where ${\bf x}=\left( \begin{array}{c} {\bf p} \\ {\bf q} \end{array} \right)
\in \R^{2N}$ and 
\begin{equation}
 \mathcal J=\left( \begin{array}{cc} 0_N & -1_N \\ 1_N & 0_N \end{array} \right)
\end{equation}
is the standard symplectic matrix, being $1_N$ the $N\times N$ identity matrix and $0_N$ the $N\times N$ matrix whose entries are all zero.

Many chaos indicators like the FLI \cite{Froeschle97}, the MEGNO \cite{Cincotta03}, the SALI \cite{Skokos01} and more recently the GALI \cite{Skokos07} require the time evolution of deviation vectors to be computed. These vectors, ${\boldsymbol \delta}(t)$, satisfy the variational equations given by
\begin{equation} 
 \dot {\boldsymbol \delta}(t) = D_{\bf x} {\bf W}{\boldsymbol \delta}(t) = \mathcal J {\bf \nabla}_{\bf x}^2 \mathcal H \,{\boldsymbol \delta}(t) \label{TM}
\end{equation}
where ${D}_{\bf x} {\bf W}$ is the Jacobian matrix of the vector field ${\bf W}$ and ${\bf \nabla}_{\bf x}^2 \mathcal H$ is the Hessian matrix of $\mathcal H$.

In \cite{Libert10}, the GSI has been introduced to simultaneously integrate both
systems of equations (\ref{HVectField}) and (\ref{TM}) by means of a
symplectic integrator. This method assumes that $\mathcal H$ may be split into
two separately integrable parts. For instance when
\begin{equation}
 \mathcal H({\mathbf p} , {\mathbf q})=A({\mathbf p})+ B({\mathbf q})\, , \label{splitting}
\end{equation}
the variational equations (\ref{TM}) can be written as
\begin{equation}
 \left( \begin{array}{c} \dot {\boldsymbol \delta}_{\mathbf p} \\ \\ \dot {\boldsymbol \delta}_{\mathbf q} \end{array} \right)=
 \left( \begin{array}{cc} 0 & -{\bf \nabla}_{\bf q^2}^2 B \\ & \\ {\bf \nabla}_{\bf p^2}^2 A & 0 \end{array} \right)\left( \begin{array}{c} {\boldsymbol \delta}_{\mathbf p} \\  \\{\boldsymbol \delta}_{\mathbf q} \end{array} \right)=
 \left( \begin{array}{c}  -{\bf \nabla}_{\bf q^2}^2 B \, \delta_{\mathbf q} \\ \\ {\bf \nabla}_{\bf p^2}^2 A \, \delta_{\mathbf p} \end{array} \right)
 =  \left( \begin{array}{c}  -{\bf \nabla}_{\boldsymbol \delta_{\bf q}} \mathcal B \\ \\{\bf \nabla}_{\boldsymbol \delta_{\bf p}} \mathcal A \end{array} \right), \label{dev:TM}
\end{equation} 
while the associated Hamiltonian $\mathcal K$ is expressed as
\begin{equation}
 \mathcal K({\mathbf p} , {\mathbf q},{\boldsymbol \delta}_{\mathbf p},{\boldsymbol \delta}_{\mathbf q}) = 
  \frac12 {\boldsymbol \delta}_{\mathbf p}^T{\bf \nabla}_{\bf p^2}^2 A {\boldsymbol \delta}_{\mathbf p} + \frac12{\boldsymbol \delta}_{\mathbf q}^T{\bf \nabla}_{\bf q^2}^2 B{\boldsymbol \delta}_{\mathbf q} =
 \mathcal A({\mathbf p},{\boldsymbol \delta}_{\mathbf p}) + \mathcal B({\mathbf q},{\boldsymbol
   \delta}_{\mathbf q})\, .
 \end{equation}
The above introduced notations will be use throughout the paper.

\subsection{Integrator}\label{SectionInt}
In the present work, two symplectic integrators have been tested. Both are based on the Campbell-Baker-Hausdorff (CBH)~\cite{Bourb1972} formula which ensures that one can find a general explicit integrator with $n$ steps of the form
\begin{equation}
S_n(\tau)=e^{c_1\tau L_A}e^{d_1\tau L_B} \cdots e^{c_n\tau L_A}e^{d_n\tau L_B}=e^{\tau L_{\mathcal G}},
\end{equation}
whose coefficients $c_i$ and $d_i$ have to be carefully chosen to get the required precision, i.e. {reach} the integrator order $m$. Integrating $\mathcal H$ at order $m$ means that we exactly evaluate $e^{\tau L_{\mathcal G}}$
where
\begin{equation}
\mathcal G=A+ B+\mathcal O(\tau^m). \label{eq:intError}
\end{equation}

In \cite{Laskar01}, four classes of symmetric symplectic integrators have been
presented: 
\begin{equation}
 \begin{array}{rcl}
 \text{SABA}_{2n}(\tau) & = & e^{c_1\tau L_A}e^{d_1\tau L_B}...e^{c_n\tau L_A}e^{d_n\tau L_B}e^{c_{n+1}\tau L_A}e^{d_n\tau L_B}e^{c_n\tau L_A}...e^{d_1\tau L_B}e^{c_1\tau L_A} \\
 \text{SABA}_{2n+1}(\tau) & = & e^{c_1\tau L_A}e^{d_1\tau L_B}...e^{c_{n+1}\tau L_A}e^{d_{n+1}\tau L_B}e^{c_{n+1}\tau L_A}...e^{d_1\tau L_B}e^{c_1\tau L_A} \\
 \text{SBAB}_{2n}(\tau) & = & e^{d_1\tau L_B}e^{c_2\tau L_A}e^{d_2\tau L_B}...e^{c_{n+1}\tau L_A}e^{d_{n+1}\tau L_B}e^{c_{n+1}\tau L_A}...e^{d_2\tau L_B}e^{c_2\tau L_A}e^{d_1\tau L_B} \\
 \text{SBAB}_{2n+1}(\tau) & = & e^{d_1\tau L_B}e^{c_2\tau
   L_A}...e^{d_{n+1}\tau L_B}e^{c_{n+2}\tau L_A}e^{d_{n+1}\tau
   L_B}...e^{c_2\tau L_A}e^{d_1\tau L_B} \, .
 \end{array}
\end{equation}
If $\varepsilon:=|B|/|A|$ is small enough, it is shown that one can find specific coefficients such that
\begin{equation}
\mathcal G=A+B+\mathcal O(\tau^{2n}\varepsilon+\tau^2\varepsilon^2). \label{eq:intErrorPert}
\end{equation}
Obviously, since the approximation error depends on the \emph{weight}
$\varepsilon$ of the perturbation, these classes of integrators are not well
suited for Hamiltonian systems that are not perturbations of integrable ones.

In general, for the Hamiltonian $\mathcal K$
associated to variational equations, the ratio $|\mathcal B|/|\mathcal A|$ is not
necessarily small. Hence, the method presented in \cite{Laskar01} is not
suitable. {To tackle this problem}, we have chosen to use another class of
symmetric and explicit symplectic integrators presented in
\cite{Yoshida90}. The latter does not take into account the importance of the
perturbation and can be defined starting from the basic bloc given by the
second order 
St\"ormer-Verlet/Leap Frog scheme, $T_{\text{2nd}}(\tau)  =
  e^{\frac12\tau L_A}e^{\tau L_B}e^{\frac12\tau L_A}$, as follows:
\begin{eqnarray}
 T_{\text{4th}}(\tau) & = & T_{\text{2nd}}\left(\frac{1}{2-2^{1/3}}\tau \right)T_{\text{2nd}}\left(-\frac{2^{1/3}}{2-2^{1/3}}\tau \right)T_{\text{2nd}}\left(\frac{1}{2-2^{1/3}}\tau \right) \\
 T_{\text{6th}}(\tau) & = & T_{\text{4th}}\left(\frac{1}{2-2^{1/5}}\tau \right)T_{\text{4th}}\left(-\frac{2^{1/5}}{2-2^{1/5}}\tau \right)T_{\text{4th}}\left(\frac{1}{2-2^{1/5}}\tau \right). 
\end{eqnarray}
In this case, the error depends only on the time step $\tau$, as in
Eq. (\ref{eq:intError}). While this method turns out to be very efficient
for the variational Hamiltonian $\mathcal K$, it does not take advantage
of the structure of ${\mathcal H}$ as a perturbation of an integrable system.  

\section{MEGNO}
\label{sec:megno}

\subsection{Definition}

According to \cite{Cincotta03}, the Mean Exponential Growth factor of Nearby Orbits is defined as
\begin{equation} 
 Y(t)=\frac 2 t \int_0^t \frac{\dot \delta(s)}{\delta(s)} s ds  \label{MEGNO}
\end{equation}
where $\delta(s)$ denotes the Euclidean norm of $\boldsymbol \delta(s)$.
A more useful and stable indicator is given by the mean MEGNO, namely
the time-average: 
\begin{equation} 
 \bar{Y}(t)=\frac 1 t \int_0^t Y(s) ds . \label{meanMEGNO}
\end{equation}

While $Y(t)$ may not converge nor admit a limit for $t\rightarrow \infty$, it
has been proven {by}~\cite{Cincotta03} that the asymptotic value of
$\bar{Y}$ provides a good characterization of the regular or chaotic nature of
orbits. Basically, $\lim_{t\rightarrow\infty} \overline Y(t)=2$ for quasi-periodic orbits on an irrational torus for a non-isochronous system
and for orbits close to stable periodic ones. In the limit case where
the orbit coincides with a stable {periodic} orbit, $\bar Y(t)$ asymptotically reaches
zero. Considering irregular orbits, $\overline Y(t)$ increases linearly
with time, being the slope half of the first Lyapunov exponent. 

\subsection{Computation}
The computation of the MEGNO and its time-average requires both integrals (\ref{MEGNO}) and (\ref{meanMEGNO}) to be solved accurately. Different methods are available. 

A straightforward approach is based on the introduction of two auxiliary functions $v$ and $w$ such that
\begin{equation} 
 v(t)=t Y(t) \quad \text{and} \quad w(t)=t \bar{Y}(t)\, , \label{vw}
\end{equation} 
whose time evolution is {directly} given by the following differential equations:
\begin{equation} 
 \dot v(t) = 2 \frac{\dot \delta(t)}{\delta(t)} t=2 \frac{\dot{\bf \delta} \cdot {\bf \delta} }{\delta^2}t \quad \text{and} \quad \dot w(t) = Y(t)= \frac{v(t)}{t}. \label{vw2}
\end{equation}
Obviously, (\ref{TM}) and (\ref{vw2}) have to be computed with the same
  integrator (see e.g. \cite{Gozdz01},\cite{Valk09}, \cite{Hinse10}). In this
case, the time step used to integrate Eq. (\ref{TM}) is fixed by the
integration of Eq. (\ref{vw2}). 
However, the use of auxiliary functions is less efficient within a
symplectic integration scheme (in fact, (\ref{vw2}) are not generally
Hamiltonian equations). Indeed, it is well-known that symplectic integrators
show very good energy conservation properties, allowing us to consider larger
step sizes and limit energy loss. In
light of these considerations, other alternatives have been studied. 

In particular, using the definition of the MEGNO for discrete time
dynamical systems (i.e. maps) given by~\cite{Cincotta03}, it has been proposed
in~\cite{Breiter05} to compute $Y(t)$ and $\bar{Y}(t)$ by observing that a
fixed step size integrator can be considered as equivalent to a discrete time
map. Hence 
\begin{eqnarray}
 Y_{\text{\tiny [Breiter et al., 2005]}}(t+h) = \frac{t}{t+h} Y_{\text{\tiny [Breiter et al., 2005]}}(t)+2\ln \frac{\delta(t+h)}{\delta(t)} \quad\text{and} \notag \\ \bar{Y}_{\text{\tiny [Breiter et al., 2005]}}(t+h) = \frac{t \bar{Y}_{\text{\tiny [Breiter et al., 2005]}}(t)+hY_{\text{\tiny [Breiter et al., 2005]}}(t+h)}{t+h}\, , 
\label{eq:breitermegno}
\end{eqnarray}
$h$ being the integration step size. Let us note that these formulas can be 
obtained by using a simple rectangular quadrature method to solve both 
integrals (\ref{MEGNO}) and (\ref{meanMEGNO}). On the other hand, a mixed scheme has
been proposed in~\cite{Gozdz03}, that relies on the computation of the
MEGNO by using the so-called trapezoidal rule~\footnote{Given a real-valued
  function $f$ and an interval $[a,b]$, one
  has~\cite{Press07} \begin{equation*} 
 \int_a^b f(x) dx = 0.5 (b-a) [f(a)+f(b)] + \mathcal
 O((b-a)^3f'')\, , \label{trap} 
\end{equation*}the second derivative being estimated on the interval $[a,b]$.}
to compute the integral in the definition~(\ref{MEGNO}) and then the discrete
time approximation for the mean MEGNO. 

In the present work, we develop further this idea: we propose to use
the trapezoidal rule to compute both integrals (\ref{MEGNO})
and~(\ref{meanMEGNO}). First, we  
rewrite~{(\ref{MEGNO})} as 
\begin{equation} 
  Y(t)=2\log \delta(t)-\frac{2}{t} \int_0^t \log{\delta(s)} ds\, ,  \label{MEGNO2}
\end{equation}
then using the trapezoidal rule we get
\begin{eqnarray}
 Y(t+h) = \frac{t}{t+h} Y(t)+ \frac{2t+h}{t+h} \ln \frac{\delta(t+h)}{\delta(t)} +\mathcal O(h^3)\label{MEGNOquad} 
\end{eqnarray}
and
\begin{eqnarray}
 \bar{Y}(t+h) = \frac{1}{t+h} \int_0^{t+h} Y(s) ds = \frac{1}{t+h} [ t \bar{Y}(t)+0.5 h (Y(t)+Y(t+h)) ]+\mathcal O(h^3)\, .  \label{meanMEGNOquad} 
\end{eqnarray}
Let us observe that the above formulas (\ref{MEGNOquad}) and (\ref{meanMEGNOquad}) improve the aforementioned ones corresponding to lower order approximations of the integrals defining MEGNO and mean MEGNO. 

In the following, Eq. (\ref{MEGNOquad}) and (\ref{meanMEGNOquad}) will be used
to compute the MEGNO whenever we use a symplectic scheme, whereas Eq. (\ref{vw2}) will be used with RK4. Let us note that in the rest of the paper we will
  be interested in asymptotic values of mean MEGNO i.e. mean
MEGNO values at the end of the integration process. In order not to introduce
{any} bias in the computation of the MEGNO and mean MEGNO, initial deviation
vectors ${\boldsymbol \delta}(0)$ will always be randomly chosen with
  uniform probability in the appropriate hypersphere.  

\section{Arnold Diffusion}
\label{sec:arndiff}

\subsection{Model}

\begin{figure}[t]
\begin{center}
\psfig{file=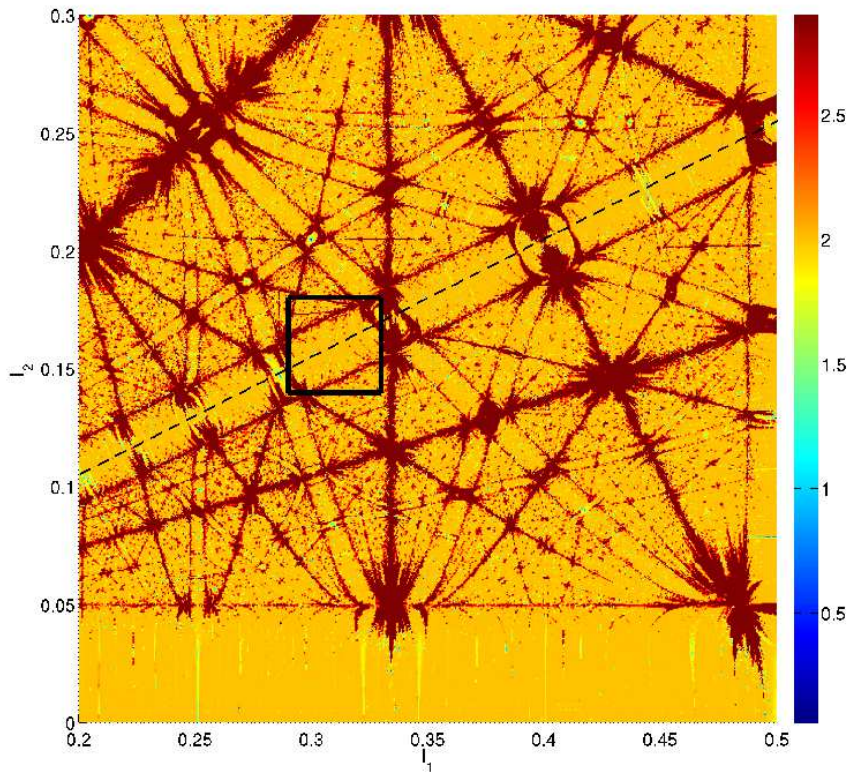,width=13cm}
\end{center}
\caption{Arnold web. Two-dimensional phase plane $(I_1,I_2)$
  represented using mean MEGNO values at $t=10^7$ time
    units (the values greater than three have been fixed to three). A
    set of 
  $600\times600$ uniformly distributed initial conditions has been integrated
  with $T_{\text{4th}}$ 
  (with time steps equal to $0.01$). Other
  initial conditions are fixed to $I_3=1$, $\phi_1=\phi_2=\phi_3=0$ and
  $\nu=0.007$. The dashed line represents the $I_1=2I_2$ resonance. The box
  encloses a part of it and is analyzed in more detail in Fig. \ref{PhaseSpaceBox}.} 
\label{PhaseSpace}
\end{figure}

As shown in \cite{Libert10}, the GSI proves to be more efficient than
non-symplectic schemes to correctly identify the behavior of a given orbit
{especially on dynamics acting on long time scales}. In this section, we will
show that this is particularly relevant in case of slow chaotic diffusion. To
that purpose, we decided to consider the following Hamiltonian system
{described} in \cite{Lega03}: 
\begin{equation}
 \mathcal H_{\text{Arnold}}(I_1,I_2,I_3,\phi_1,\phi_2,\phi_3) = \frac12(I_1^2+I_2^2)+I_3+\nu \frac{1}{cos(\phi_1)+cos(\phi_2)+cos(\phi_3)+4}, \label{Harnold}
\end{equation}
where actions $I_1,I_2,I_3\in \R$ and angles $\phi_1,\phi_2,\phi_3\in\T$ are
canonically conjugate variables and $\nu$ is
assumed to be a small parameter. 

Given the structure of
(\ref{Harnold}), $\dot \phi_1=I_1$, $\dot \phi_2=I_2$ and $\dot
\phi_3=1$. Hence, each straight line  
$$
 k_1I_1+k_2I_2+k_3=0 \; , \quad (k_1,k_2,k_3)\in\Z^3 \setminus \{0\}
$$
on the two-dimensional plane $(I_1,I_2)$ represents a resonance.

 As illustrated in Fig.~\ref{PhaseSpace}, most relevant resonances are clearly
 visible on the plane $(I_1,I_2)$, to form the so-called \emph{Arnold
   Web}. Mean MEGNO values are shown for a grid of $600\times600$ equally spaced
   initial conditions in this plane. Other initial conditions are $I_3=1$ and
 $\phi_1=\phi_2=\phi_3=0$, and the parameter $\nu$ has been fixed to $0.007$,
 as in the rest of this work. This value needs to be small in order to avoid
 resonances overlap. Besides, as pointed out in \cite{Lega03}, the smaller the
 perturbation, the slower  the diffusion. The GSI has been used with
 $T_{\text{4th}}$ integrator with a fixed time step $\tau=0.01$ and an
 integration time of $10^7$. The dashed line highlights the $I_1=2I_2$
 resonance. This web of resonances is of particular interest. Indeed, it has
 been proven (see \cite{Arnold63}) that Arnold diffusion exists along
 resonances. Moreover, in \cite{Lega03}, a numerical proof of diffusion along
 resonances for this model is given.  

\begin{figure}[t]
\begin{center}
\psfig{file=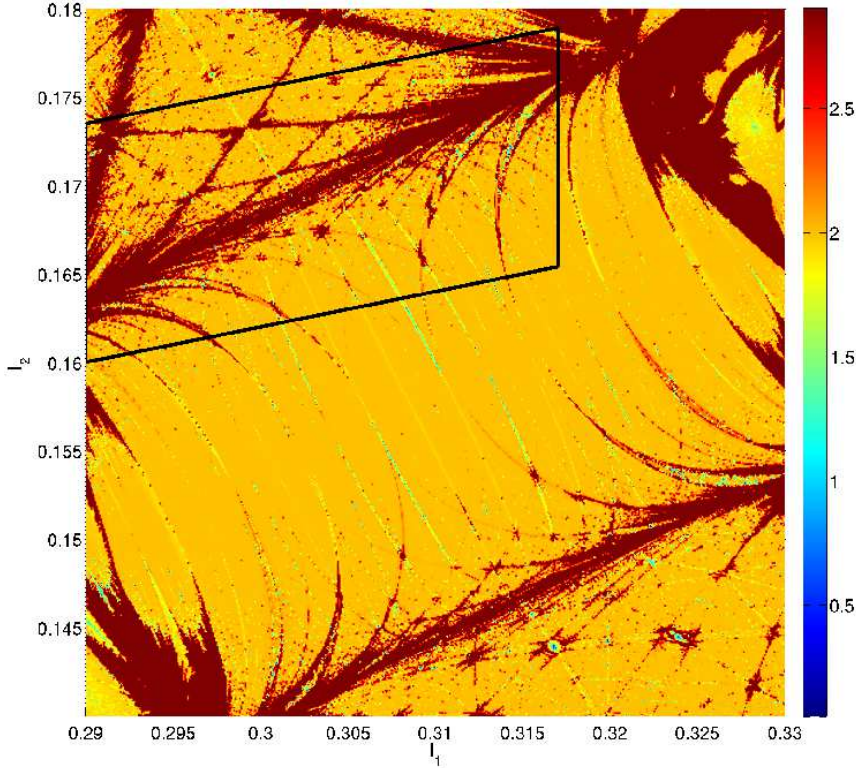,width=13cm}
\end{center}
\caption{Smaller portion of Arnold web. Two-dimensional phase plane
  $(I_1,I_2)$ represented using mean MEGNO values at $t=10^7$ time units
  (the  value{s} greater than three have been fixed to three). A set of
  $600\times600$ uniformly distributed initial conditions have been integrated with
  $T_{\text{4th}}$ (with time steps equal to $0.01$). Other initial conditions are fixed to $I_3=1$ and
  $\phi_1=\phi_2=\phi_3=0$. The parallelogram represents the region in which
  $100$ initial conditions have been considered to compare the GSI to a
    non symplectic integration scheme.} 
\label{PhaseSpaceBox}
\end{figure} 

The analysis presented in Section \ref{ArnoldAnalysis} will be performed in
the region delimited by $0.29\leq I_1\leq0.33$ and $0.14\leq I_2\leq0.18$,
centered on the $I_1=2I_2$ resonance. This small region is enclosed in the
  box shown in Fig. \ref{PhaseSpace}. An enlargement of
  this box is presented in Fig. \ref{PhaseSpaceBox}. Again, a grid of
  $600\times600$ equally spaced initial conditions has been numerically
  integrated using the GSI with  $T_{\text{4th}}$ integrator up to $10^7$
    time units. Other initial conditions and parameters are the same as the
  ones used to produce Fig. \ref{PhaseSpace}. In the following analysis, we
  will consider several orbits around the top hyperbolic border (in brown in
  Fig. \ref{PhaseSpaceBox}) of the resonance where diffusion is actually
  confined. 

\subsection{Analysis}\label{ArnoldAnalysis}

In this section, we will compare the results on the correct determination of
regular or chaotic orbits behavior obtained with the GSI and a
non-symplectic scheme. Through an analysis of the maximum relative errors on
the energy, percentage of correctly identified orbits and CPU time, we will
show that our symplectic scheme outperforms the non-symplectic one in
the determination of the behavior of orbits. 

\begin{figure}[b]
\begin{center}
\psfig{file=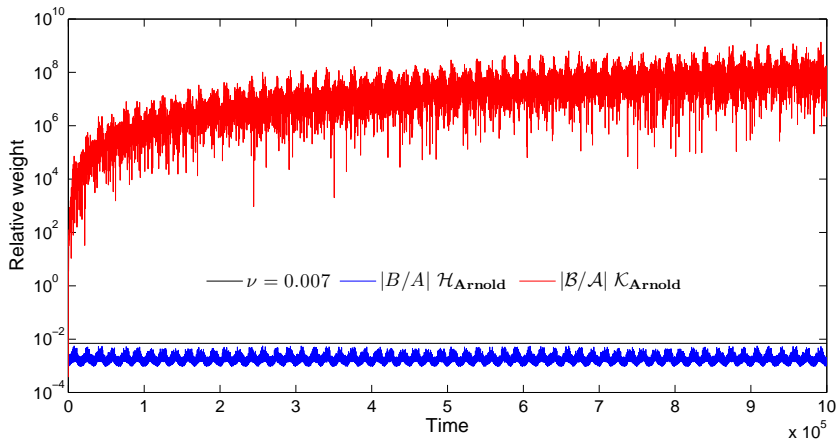,width=13cm} 
\end{center}
\caption{Relative weight of $A$ and $B$ parts (respectively $\mathcal A$ and
  $\mathcal B$), 
  for both Hamiltonian systems $\mathcal H_{\text{Arnold}}$ and $\mathcal
  K_{\text{Arnold}}$. One generic orbit and the associated
  variational equations have been integrated with $T_{\text{4th}}$ with a time
  step $\tau=0.01$. The horizontal line corresponds to $\nu=0.007$.}
\label{Perturbation}
\end{figure}

As it is necessary to consider long integration times and a lot of different
initial conditions, it has been decided to use and compare fourth-order
integrators. In fact, this turns out to be a good compromise between
reliability of the numerical results, as measured in term of relative energy
loss, and number of evaluations of the vector fields, translated easily into
required CPU time. On the one hand, we considered the well-known
non-symplectic fourth-order Runge Kutta (RK4) integrator. While it is simple
to implement, it is also robust and efficient. On the other hand, both
symplectic integrators presented in Section \ref{SectionInt} have been tested.  
As far as only the Hamiltonian system (\ref{Harnold}) is concerned, SABA and
  SBAB integrator classes outperform Yoshida integrator. Indeed, given that
  $\nu$ is small, this system can be seen as a perturbation of an integrable
  system. Hence, the error (\ref{eq:intErrorPert}) is smaller than Yoshida's
  one (\ref{eq:intError}). However, SABA and SBAB classes performances become
  rapidly poor for 
  the Hamiltonian system $\mathcal K_{\text{Arnold}}$ associated to the
  variational equations. It is mainly due to the structure of $\mathcal
  K_{\text{Arnold}}$. As time increases, the weight of the $\mathcal{B}$ part, say the
  perturbation, of the Hamiltonian $\mathcal{K}_{\text{Arnold}}$ may become
 larger than the $\mathcal{A}$ part, or vice versa without any possible
 control, see Fig.~\ref{Perturbation}. This implies that the 
  error on the energy increases too. Fortunately, Yoshida integrators do not
  depend 
  on this kind of consideration and are then more suited for this
  application. As already stated in Section \ref{SectionInt}, Yoshida
  integrators do not take advantage of the structure of $\mathcal
  H_{\text{Arnold}}$ but it is conversely useful for the integration of
  variational equations. Hence, for the purpose of the present study, it has
  been decided to compare RK4 integrator to $T_{\text{2nd}}$ and
  $T_{\text{4th}}$.

The comparison has been performed considering $100$ orbits whose initial
conditions ($I_1(0),I_2(0)$) are uniformly distributed around the top
hyperbolic border of the resonance (the parallelogram in
Fig. \ref{PhaseSpaceBox}). First, for each one of these orbits, a
\emph{reference} value of the mean MEGNO has been computed with
$T_{\text{4th}}$, a time step equal to $\tau=0.01$ and an integration time of
$2.5\times10^6$. Afterwards, for time steps between $0.01$ and $1$, the same orbits
have been numerically integrated with both methods (the GSI and the
non-symplectic scheme) in order to obtain the corresponding
MEGNO values. Eventually, a comparison with the \emph{reference} MEGNO values
has been done, enabling us to classify orbits as correctly identified or
not. At the same time, CPU times and maximum relative errors in energy have
been stored. In order to reduce numerical truncation errors, it has been
  decided to achieve this study using quadruple precision.  

A relevant indicator to test the goodness of the numerical integration,
  is to compare the maximum relative errors in energy
\begin{equation}
 \Delta E/E=\max_{\stackrel{0\leq t\leq 2.5\times10^6}{\tiny
     i=1,...,100}}|E_i(t)-E_i(0)|/|E_i(0)|\, ,
\end{equation}
$E_i(t)$ being the energy, i.e. the value of the Hamilton function, at
time $t$ on the $i$--th orbit.

A second important indicator, related to the speed of the
integration algorithm, is the CPU time
\begin{equation}
T_{\text{CPU}}=\sum_{i=1}^{100} T_{\text{CPU}}^i\, ,
\end{equation}
$T_{\text{CPU}}^i$ being the CPU time needed to integrate the $i$--th
  orbit and the associated deviation vector on the defined time span.

\begin{figure}[b]
\begin{center}
\hspace*{-0.625cm}\psfig{file=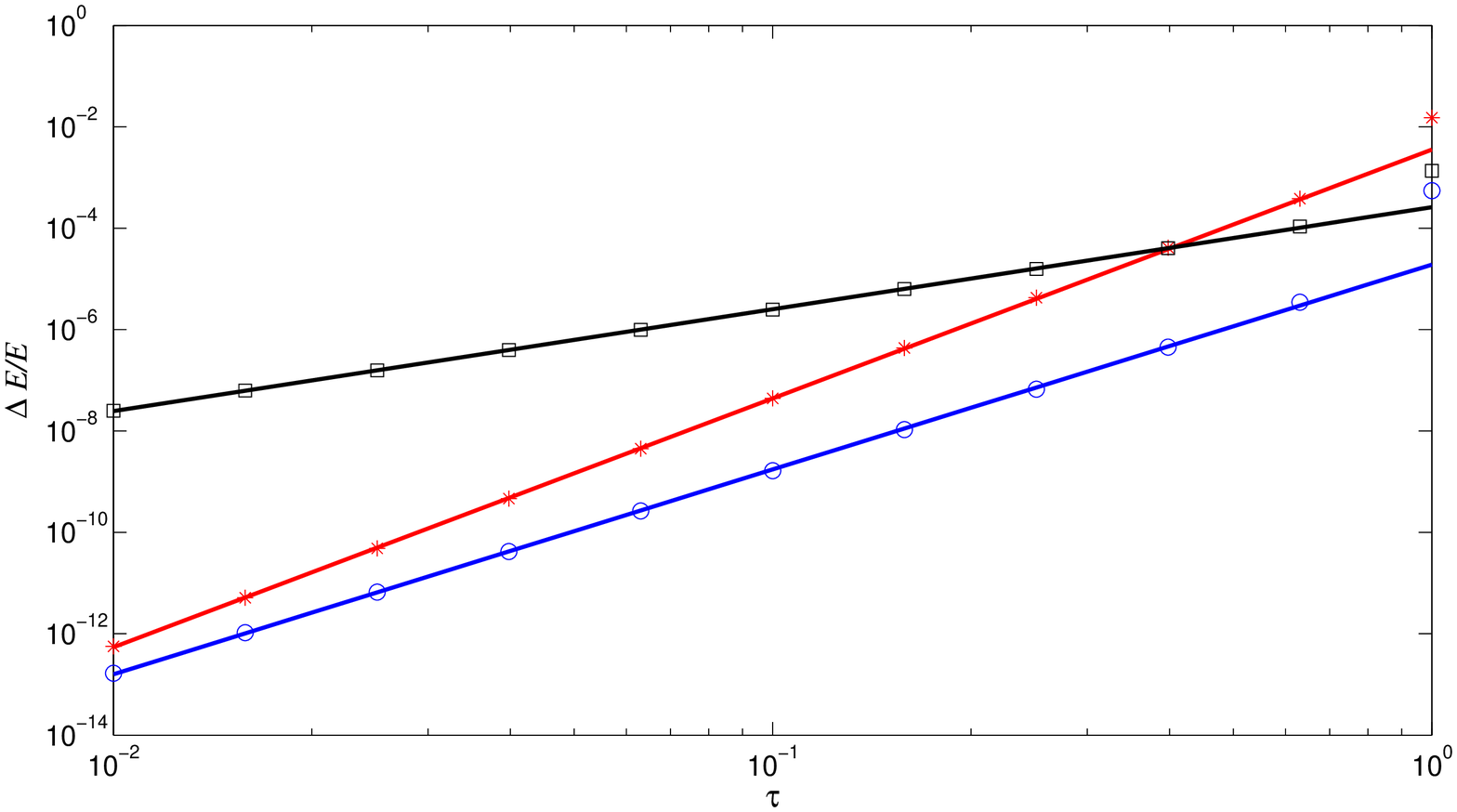,scale=0.45}
\hspace*{-1.2cm}\psfig{file=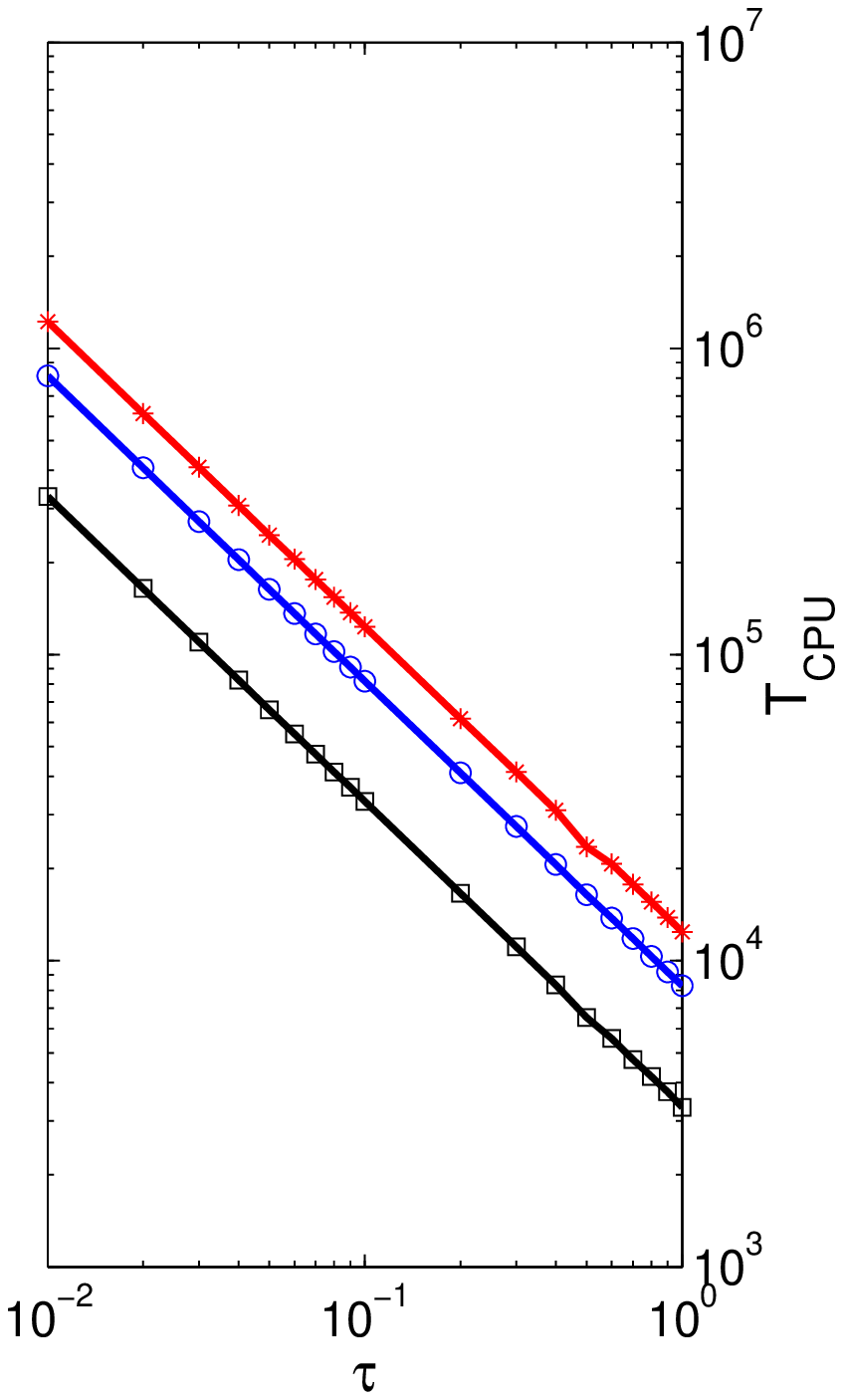,scale=0.45}
\end{center}
\caption{Maximum relative errors in energy $\Delta E/E$ (left panel) and CPU
  time $T_{\text{CPU}}$ (right panel) as a function of the time step, in
  logarithmic 
  scale. The integration time for this analysis has been set to
  $2.5\times 10^6$ time units. The comparison involved $100$ orbits whose
  initial 
  conditions $\left(I_1(0),I_2(0)\right)$ have been taken around the top
  hyperbolic border of the $I_1=2I_2$ resonance (see
  Fig.~\ref{PhaseSpaceBox}). Symbols 
  are : (red) $*$ RK4 integrator, (blue) $\bigcirc$ the 4th order Yoshida
  integrator and (black) $\square$ the 2nd order Yoshida integrator. Straight
  lines denote best linear fits (left panel).}
\label{Energy}
\end{figure}

Both indicators are reported in Fig. \ref{Energy} as functions of the
time step for the different integrators. It appears that $T_{\text{4th}}$
shows always smaller energy loss than RK4 integrator. Moreover, as time step
increases, this difference becomes larger too. Also note that the maximum
error becomes larger with RK4 than with $T_{\text{2nd}}$
beyond $\tau\simeq 
0.25$. That means that, even if $T_{\text{2nd}}$ is only a second order
integrator, it is more reliable than RK4 when using big time steps. Another
advantage of the GSI is the relatively low required CPU time
($T_{\text{CPU}}$) in comparison to RK4. This is particularly important as we
are considering lots of different initial conditions and long integration
times. Obviously, the lower-order $T_{\text{2nd}}$ asks less CPU time than
$T_{\text{4th}}$. 

\begin{figure}[t]
\begin{center}
\psfig{file=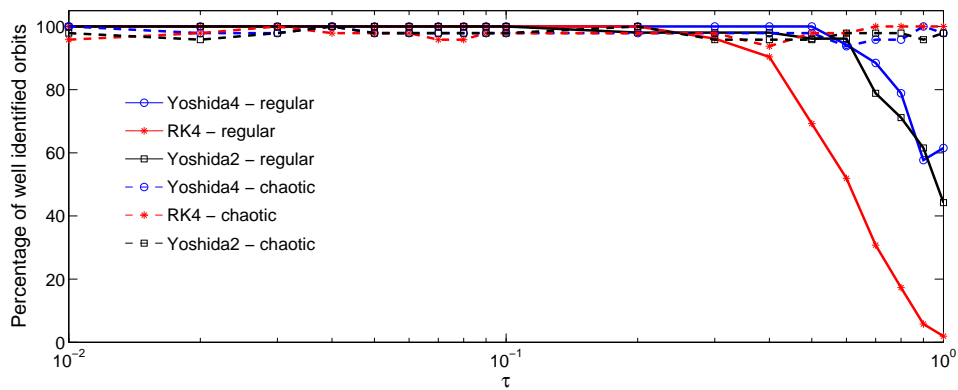,width=15cm}
\end{center}
\caption{Percentages of correctly identified orbits with respect to time step
  in logarithmic scale. The comparison involve $100$ orbits whose
  initial conditions $\left(I_1(0),I_2(0)\right)$ have been taken around the
  top hyperbolic border of the $I_1=2I_2$ resonance (see
  Fig.~\ref{PhaseSpaceBox}). Solid lines and dashed lines represent
  respectively the identification of regular orbits and chaotic orbits.} 
\label{Perc1}
\end{figure}

Moreover, the GSI correctly identifies more orbits than RK4 as time steps
  increase (see Fig. \ref{Perc1}). Indeed, mean MEGNO values computed by
means of RK4 are wrong for regular orbits when the time step is greater than
$0.1$. The percentage of well identified regular orbits even reaches zero
while, at the same time, $T_{\text{4th}}$ is still beyond $50\%$. This
difference is less discernible for chaotic orbits, since a small drift from
the orbit and/or tangent direction does not lead to completely different
behaviors. However, in the following, we will use the total percentage, $p$,
  that presents a summary of both results. Eventually, let us point out that
  the lower-order $T_{\text{2nd}}$ integrator manages to identify correctly
  approximately the same percentage of orbits as $T_{\text{4th}}$. 

\begin{figure}[t]
\begin{center}
\psfig{file=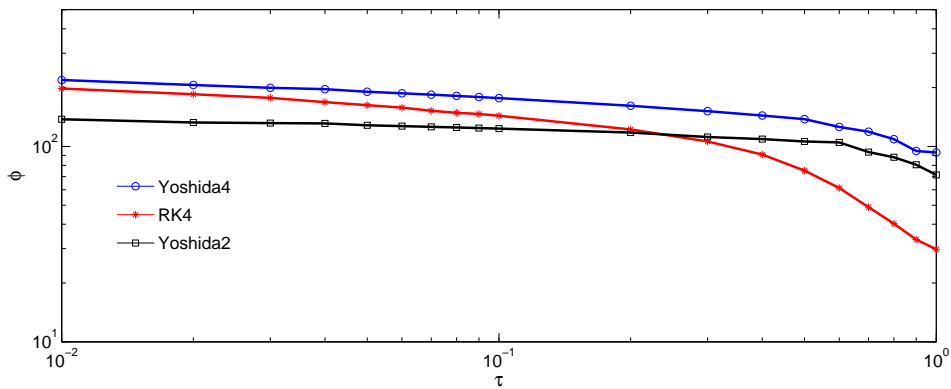,width=15cm}
\end{center}
\caption{Efficiency
  index $\phi$ with respect to time step. The comparison involved $100$ orbits
  whose 
  initial conditions $(I_1(0),I_2(0))$ have been taken around the top hyperbolic
  border of the $I_1=2I_2$ resonance.} 
\label{fig:index}
\end{figure}

Eventually, all these observations can be recombined into a single
efficiency index. As introduced in \cite{Libert10}, 
\begin{equation}
 \phi=p |\log_{10}(\Delta E/E)||\log_{10}(T_{\text{CPU}})|^{-1}
\label{eq:effindex}
\end{equation}
enables us to quantify the \emph{efficiency} of each method. The larger
$\phi$, the better the method. On the one hand, the percentage of correctly
identified orbits must be as big as possible. On the other hand,
relative error in energy ($\in[0,1]$) and CPU time must remain low. 
The evolution of $\phi$ with respect to time step is shown in
  Fig. \ref{fig:index}. It turns out that this index presents larger values
  for computations realized with the GSI, coupled to $T_{\text{4th}}$. It
  results obviously from previous considerations. For relatively small times
  steps ($\tau<0.02$), the non-symplectic scheme (RK4) behaves similarly to
  the GSI. After that, this method is quickly penalized by its energy loss and
  lower percentage of well identified orbits. Eventually, efficiency index for
  the GSI used with $T_{\text{2nd}}$ presents similar
  behavior to the ones of $T_{\text{4th}}$ but on a lower
  level. It comes directly from its larger maximum relative error in
  energy. However, at time step $\tau\simeq0.2$, the
  GSI with $T_{\text{2nd}}$ becomes more efficient than the non-symplectic
  method, due to the joint effect of better energy conservation and number of
  correctly estimated orbits.

Let us observe that the
same analysis has been performed in double precision. Results concerning the
percentage of correctly identified orbits are exactly the same. Obviously, CPU
times are proportionally smaller than the present ones. The main difference
occurs with the evolution of maximum relative energy errors with respect to
time step which are not perfect straight lines anymore. In particular,
numerical truncation errors appear for too small relative errors in
energy. This leads to an inaccurate efficiency index. 

\section{Conclusion}
\label{sec:concl}

In this work, we have shown that the GSI is a powerful tool to characterize
  the regular or chaotic behavior of orbits in Hamiltonian systems. It proves
  to be especially efficient for dynamics acting on long time scales like the
  Arnold diffusion problem analyzed here, as energy conservation properties of
  symplectic integration schemes are fully pointed up. 

As it only asks for the knowledge of a single deviation vector, the MEGNO
  indicator is less time consuming than other chaos indicators and should be
  used jointly with the GSI. To numerically compute both MEGNO and mean MEGNO,
  we have introduced a quadrature method based on the trapezoidal rule. 

Furthermore, it has been shown that a particular attention has to be paid to
  the choice of the symplectic integrator. One has to bear in mind that both
  Hamiltonian systems associated to the equations of motion and the
  variational equations can present very different structures. For this
  reason, we argue for the use of Yoshida's symplectic integrator. 

Finally, our analysis shows that the GSI outperforms non-symplectic
  schemes. Indeed, large time steps are allowed, smaller computation times are
  needed and the energy loss is more limited. For all these considerations, we
  claim that the GSI method with the MEGNO indicator and Yoshida integrator
  turns out to be a reliable and time sparing method to correctly determine
  the behavior of orbits associated to Hamiltonian systems. 

\nonumsection{Acknowledgments}

The work of Ch. Hubaux is supported by an FNRS Research Fellowship. The work
of A.-S. Libert is supported by an FNRS Postdoctoral Research
Fellowship. Numerical simulations were made on the local computing resources
(Cluster URBM-SYSDYN) at the University of Namur (FUNDP, Belgium).

\end{document}